\documentclass[12pt]{article}

\usepackage[breaklinks,hypertexnames=false]{vv}  
\usepackage{amsmath}
\usepackage{sariel,wide}
\usepackage{amssymb}
\usepackage{algorithm, algorithmic}
\usepackage{euscript}
\usepackage{graphicx}

\newcommand{\com}[1]{}

\newcommand{\ConstA}{\ensuremath{c_1}}

\newcommand{\ConstReach}{c_{\mathrm{reach}}}
\newcommand{\ConstGap}{c_g}
\newcommand{\ConstGapBlowup}{c_{gbu}}
\newcommand{\ConstLower}{c_{low}}
\newcommand{\ConstLowerFinal}{c_{low}'}
\newcommand{\ConstUpper}{c_{up}}
\newcommand{\ConstLfsShrink}{c_{shrink}}

\newcommand{\Size}{\mathrm{size}}
\newcommand{\lfs}{\ensuremath{{\mathrm{lfs}}}}
\newcommand{\ConstLifeSpan}{c_{\mathrm{span}}}
\newcommand{\Parent}{\mathrm{parent}}

\newcommand{\Pair}[2]{(#1,#2)}
\newcommand{\PFinal}{\EuScript{F}}

\newcommand{\ConstReachExpr}{2 \ConstUpper \ConstGapBlowup}

\newcommand{\OffCenter}{\text{off-center}}

\newcommand{\Heap}{\mathcal{HP}}
\def\QT{\mathcal{QT}}

\def\extractMin{\tt{extractMin}}

\providecommand{\tildegen}{{\protect\raisebox{-0.1cm}
      {\symbol{'176}\hspace{-0.03cm}}}}

\newcommand{\Leaf}{\mathbf{d}}
\newcommand{\LeftTurn}{\ensuremath{\mathtt{left\_turn}}}
\newcommand{\RightTurn}{\ensuremath{\mathtt{right\_turn}}}
\newcommand{\Crescent}{\ensuremath{\mathrm{crescent}}}

\newcommand{\gap}{\mathrm{gap}}
\providecommand{\Ungor}{\"{U}ng\"{o}r}
\newcommand{\ACT}[1]{{\overline{#1}}}
\newcommand{\Cell}{\Box}
\newcommand{\CellA}{{\Box}'}

\newcommand{\AlgRefine}{{\tt{Delaunay\_Refinement}}}

\newcommand{\LevelInd}{\mathsf{t}}
\newcommand{\TwoFigures}[6]{\noindent\begin{tabular}{cc}
        \begin{minipage}{0.48\linewidth}
            \centering {{\includegraphics[#2]{{#1}}}}
        \end{minipage}
        \vspace*{0.05cm}
        &
        \begin{minipage}{0.48\linewidth}
            \centering {\includegraphics[#5]{{#4}}}
        \end{minipage}
        \vspace*{0.05cm}
        \\
        {#3} & {#6}
    \end{tabular}
    \vspace*{-0.45cm} }

\providecommand{\SarielThanks}[1]{\thanks{Department of Computer
      Science; 
      University of Illinois; 
      201 N. Goodwin Avenue;
      Urbana, IL, 61801, USA;
      {{\tt{sariel\string@uiuc.edu}}}; {\tt
         \url{http://www.uiuc.edu/\string~sariel/}.} #1}}
\providecommand{\AlperThanks}[1]{\thanks{Department of Computer and Information
      Science and Engineering, University of Florida, Gainesville, FL
      32605; USA; {\url{http://www.cise.ufl.edu/\string~ungor/}};
      {\tt{ungor\string@cise.ufl.edu}}. #1}}

\title{A Time-Optimal Delaunay Refinement Algorithm in Two
   Dimensions\thanks{See \urlSarielPaperExt{04/opt\string_del/}
      {04/opt\_del/} for the most recent version of this paper.}}

\author{Sariel Har-Peled\SarielThanks{Work on this paper was partially
      supported by a NSF CAREER award CCR-0132901.  }%
   \and%
   Alper \"{U}ng\"{o}r\AlperThanks{This work was initiated
      during the second author's postdoctoral studies at Duke
      University and was partially supported by NSF under the ITR
      grant CCR-00-86013.}}

\begin{document}


\maketitle

\begin{abstract}
    We propose a new refinement algorithm to generate size-optimal
    quality-guaranteed Delaunay triangulations in the plane. The
    algorithm takes $O(n \log n + m)$ time, where $n$ is the input
    size and $m$ is the output size.  This is the first time-optimal
    Delaunay refinement algorithm.
\end{abstract}



\section{Introduction}
\seclab{introduction}

Geometric domain discretizations (i.e., meshing) are essential for
computer-based simulations and modeling.  It is important to avoid
small (and also very large) angles in such discretizations in order to
reduce numerical and interpolation errors \cite{sf-afem-73}.  Delaunay
triangulation maximizes the smallest angle among all possible
triangulations of a given input and hence is a powerful
discretization tool. Depending on the input configuration, however,
Delaunay triangulation can have arbitrarily small angles.  Thus,
Delaunay refinement algorithms which iteratively insert additional
points were developed to remedy this problem.  There are other domain
discretization algorithms including the quadtree-based algorithms
\cite{beg-pgmg-94,mv-qmghd-00} and the advancing front algorithms
\cite{l-pggaf-96}.  Nevertheless, Delaunay refinement method is
arguably the most popular due to its theoretical guarantee and
performance in practice.  Many versions of the Delaunay refinement is
suggested in the literature \cite{c-gqtm-89, eg-simi-01, m-tedra-04,
   mpw-wwraw-03, r-nsaqt-93, s-drmg-97,u-ocnts-04}.
 
The first step of a Delaunay refinement algorithm is the construction
of a constrained or conforming Delaunay triangulation of the input
domain.  This initial Delaunay triangulation is likely to have bad
elements.  Delaunay refinement then iteratively adds new points to the
domain to improve the quality of the mesh and to ensure that the mesh
conforms to the boundary of the input domain.  The points inserted by
the Delaunay refinement are \emph{Steiner points}.  A sequential
Delaunay refinement algorithm typically adds one new vertex at each
iteration.  Each new vertex is chosen from a set of candidates --- the
circumcenters of bad triangles (to improve mesh quality) and the
mid-points of input segments (to conform to the domain boundary).
Chew \cite{c-gqtm-89} showed that Delaunay refinement can be used to
produce quality-guaranteed triangulations in two dimensions.  Ruppert
\cite{r-nsaqt-93} extended the technique for computing not only
quality-guaranteed but also size-optimal triangulations.  Later,
efficient implementations \cite{s-drmg-97}, extensions to three
dimensions \cite{dbs-gttd-92,s-drmg-97}, generalization of input type
\cite{s-drmg-97,mpw-wwraw-03}, and parallelization of the algorithm
\cite{stu-pdraa-02} were also studied.

\begin{figure}[t]
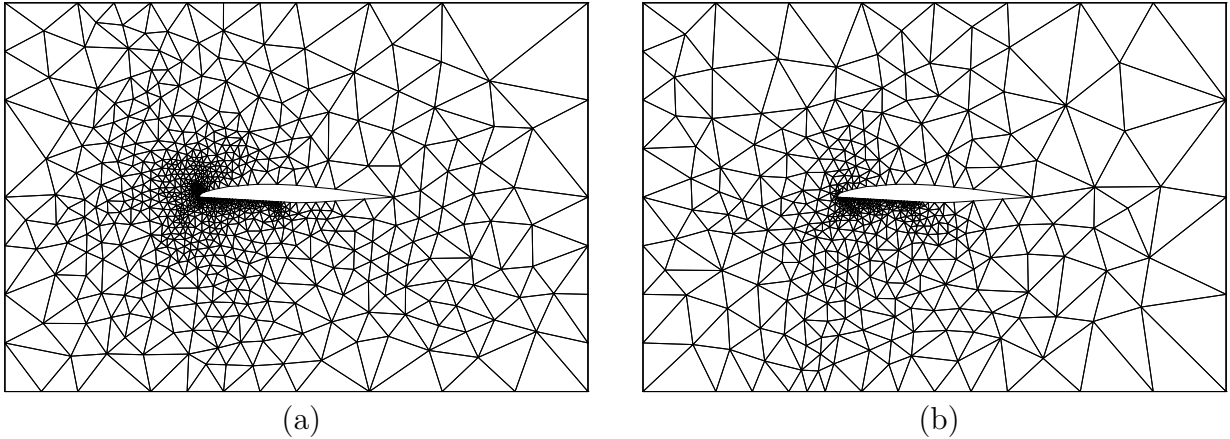

    \TwoFigures{figs/bcc}{angle=90,width=.99\textwidth}{(a)}
    {figs/boc}{angle=90,width=.99\textwidth}{(b)}

    \caption{Circumcenter vs. off-center insertion on an airfoil model.
       Smallest angle in both meshes is $32^\circ$. Delaunay
       refinement with circumcenters inserts 731 Steiner points and
       results in a mesh with 1430 triangles (a). On the other hand,
       Delaunay refinement with off-centers inserts 441 points and
       generates a mesh with 854 triangles (b).}
    \figlab{airfoil}
\end{figure}

Recently, the second author proposed a new insertion strategy for
Delaunay refinement algorithm \cite{u-ocnts-04}.  He introduced the
so-called \emph{off-centers} as an alternative to circumcenters.
Off-center of a bad triangle, like circumcenter, is on the bisector of
the shortest edge. However, for relatively skinny triangles it is
closer to the shortest edge than the circumcenter is. It is chosen
such that the triangle formed by the endpoints of the shortest edge
and the off-center is barely of good quality. Namely, the off-center
insertion is a more ``local'' operation in the mesh than circumcenter
insertion.  It is shown in \cite{u-ocnts-04} that this new Delaunay
refinement algorithm has the same theoretical guarantees as the
Ruppert's refinement, and hence, generates quality-guaranteed
size-optimal meshes.  Moreover, experimental study indicates that
Delaunay refinement algorithm with off-centers inserts considerably
fewer Steiner points than the circumcenter insertion algorithms and
results in smaller meshes. For instance, when the smallest angle is
required to be $32^\circ$, the new algorithm inserts about 40\% less
points and outputs a mesh with about 40\% less triangles (see
\figref{airfoil}). This implies substantial reduction not only in mesh
generation time, but also in the running time of the application
algorithm.  This new off-center based Delaunay refinement algorithm is
included in the fifth release of the popular
Triangle\footnote{\url{http://www-2.cs.cmu.edu/\string~quake/triangle.html}}
software. Shewchuk observed (personal communication) in this new
implementation, that unlike circumcenters, computing off-centers is
numerically stable.

Original Delaunay refinement algorithm has quadratic time complexity
\cite{r-nsaqt-93}.  This compares poorly to the time-optimal quadtree
refinement algorithm of Bern \etal{} \cite{beg-pgmg-94} which runs in
$O(n \log n + m)$ time, where $m$ is the minimum size of a good
quality mesh.  The first improvement was given by Spielman \etal{}
\cite{stu-pdraa-02} as a consequence of their parallelization of the
Delaunay refinement algorithm.  Their algorithm runs in $O(m \log m
\log^2(L/h))$ time (on a single processor), where $L$ is the diameter
of the domain and $h$ is the smallest feature in the input.  Recently,
Miller \cite{m-tedra-04} further improved this describing a new
sequential Delaunay refinement algorithm with running time $O((n
\log(L/h) + m)\log m)$.  In this paper, we present the first time
optimal Delaunay refinement algorithm. As Steiner points, we employ
off-centers and generate the same output as in \cite{u-ocnts-04}.  Our
improvement relies on avoiding the potentially expensive maintenance
of the entire Delaunay triangulation. In particular, we avoid
computing very skinny Delaunay triangles, and instead we use a
scaffold quadtree structure to efficiently compute, locate and insert
the off-center points.  Since the new algorithm generates the same
output as the off-center based Delaunay refinement algorithm given by
\Ungor{} \cite{u-ocnts-04}, it is still a Delaunay refinement
algorithm.  In fact, our algorithm implicitly computes the relevant
portions of the Delaunay triangulation.

The rest of the paper is organized as follows: In \secref{pre} we
survey the necessary background.  In \secref{loose:pairs}, we formally
define the notion of loose pairs to identify the points that
contribute to bad triangles in a Delaunay triangulation.  Next, we
describe a simple (but not efficient) refinement algorithm based on
iterative removal of loose pairs of points.  In \secref{new:alg}, we
describe the new time-optimal algorithm and prove its correctness.  We
conclude with directions for future research in \secref{conclusions}.


\section{Background}
\seclab{pre}

In two dimensions, the input domain $\Omega $ is usually represented
as a \emph{planar straight line graph}\index{planar straight line
   graph (PSLG)} (PSLG) --- a proper planar drawing in which each edge
is mapped to a straight line segment between its two endpoints
\cite{r-nsaqt-93}.  The segments express the \emph{boundaries} of
$\Omega $ and the endpoints are the \emph{vertices} of $\Omega $.  The
vertices and boundary segments of $\Omega $ will be referred to as the
input \emph{features}.  A vertex is incident to a segment if it is one
of the endpoints of the segment.  Two segments are incident if they
share a common vertex.  In general, if the domain is given as a
collection of vertices only, then the boundary of its convex hull is
taken to be the boundary of the input.

The \emph{diametral circle} of a segment is the circle whose diameter
is the segment.  A point is said to \emph{encroach} a segment if it is
inside the segment's diametral circle.

Given a domain $\Omega$ embedded in $\Re^2$, the \emph{local feature
   size} of each point $x \in \Re^2$, denoted by $\lfs_{\Omega} (x)$,
is the radius of the smallest disk centered at $x$ that touches two
non-incident input features.  This function is proven
\cite{r-nsaqt-93} to have the so-called \emph{Lipschitz property},
i.e., $\lfs_{\Omega}(x) \le \lfs_{\Omega}(y) + \dist{xy}$, for any two
points $x,y \in \Re^2$.

In this extended abstract, we concentrate on the case where $\Omega$
is a set of points in the plane contained in the square $[1/3,2/3]^2$.
We denote by $P$ the current point set maintained by the refinement
algorithm, and by $\PFinal$ the final point set generated.

Let $P$ be a point set in $\Re^d$.  A simplex $\tau$ formed by a
subset of $P$ points is a \emph{Delaunay simplex} \index{Delaunay
   simplex} if there exists a circumsphere of $\tau $ whose interior
does not contain any points in $P$.  This empty sphere property is
often referred to as the \emph{Delaunay property}.  The {Delaunay
   triangulation} \index{Delaunay triangulation} of $P$, denoted
$Del(P)$, is a collection of all Delaunay simplices.  If the points
are in general position, that is, if no $d+2$ points in $P$ are
co-spherical, then $Del(P)$ is a simplicial complex.  The Delaunay
triangulation of a point set can be constructed in $O (n\log n)$ time
in two dimensions \cite{e-gtmg-01}.

In the design and analysis of the Delaunay refinement algorithms, a
common assumption made for the input PSLG is that the input segments
do not meet at junctions with small angles.  Ruppert \cite{r-nsaqt-93}
assumed, for instance, that the smallest angle between any two
incident input segment is at least $90^\circ$.  A typical Delaunay
refinement algorithm may start with the \emph{constrained Delaunay
   triangulation} \cite{c-cdt-89} of the input vertices and segments
or the Delaunay triangulation of the input vertices.  In the latter
case, the algorithm first splits the segments that are encroached by
the other input features.  Alternatively, for simplicity, we can
assume that no input segment is encroached by other input features. A
preprocessing algorithm, which is also parallizable, to achieve this
assumption is given in \cite{stu-pdraa-02}.

For technical reasons, as in \cite{r-nsaqt-93}, we put the input
$\Omega$ inside a square $B$.\footnote{In fact the reader might find
   it easier to read the paper, by first ignoring the boundary, (e.g.,
   considering the input is a periodic point set).}  This is to avoid
growth of the mesh region and insertion of infinitely many Steiner
points. Let $MB = [1/3,2/3]^2$ be the minimum enclosing square of
$\Omega$. The side length of $B=[0,1]^2$ is three times that of $MB$.
We insert points on the edges of $B$ to split each into three.  This
guarantees that no circumcenter falls outside $B$.  We maintain this
property throughout the algorithm execution by refining the boundary
edges as necessary.

\begin{figure}[t]
    \centerline{\includegraphics[width=.4\textwidth]{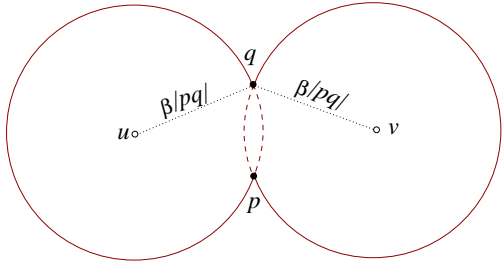}}

    \caption{Flower of a pair of points $p$ and $q$.}
    \figlab{flower}
\end{figure}

\emph{Radius-edge ratio} of a triangle is the ratio of its
circumradius to the length of its shortest side.  A triangle is
considered \emph{bad} if its radius-edge ratio is larger than a
pre-specified constant $\beta \geq \sqrt{2}$.  This quality measure is
equivalent to other well-known quality measures, such as smallest
angle and aspect ratio in two dimensions \cite{r-nsaqt-93}. Consider a
bad triangle, and observe that it must have an angle smaller or equal
to $\alpha$, where $\alpha = \arcsin( 1/2\beta )$

\tabref{notations} (in the appendix) contain a summary of the notation
used in this paper.

\section{Loose Pairs vs.~Bad Triangles}
\seclab{loose:pairs}

In the following, we use $\beta$ to denote the user specified constant
for radius-edge ratio threshold. Accordingly, $\alpha$ denotes the
threshold for small angles.

\begin{defn}
    For a pair of vertices $p$ and $q$, let $\Leaf_l(p,q)$ be the disk
    with center $u$ such that $pqu$ is a \LeftTurn{} and $\dist{up} =
    \dist{u q} = \beta \dist{p q}$.  Similarly, let $\Leaf_r(p,q)$ be
    the disk with center $v$ such that $pqv$ is a \RightTurn{} and
    $\dist{v p} = \dist{v q} = \beta \dist{p q}$.  We call the union
    of the disks $\Leaf_l(p,q)$ and $\Leaf_r(p,q)$ the \emph{flower of
       $pq$}. Moreover, $\Leaf_l(p,q)$ is called the \emph{left leaf}
    of the flower and $\Leaf_r(p,q)$ is called the \emph{right leaf}
    of the flower.
\end{defn}

A pair of vertices $\Pair{p}{q}$ in $P$ is a \emph{loose pair} if
either the left or the right leaf of the flower of $pq$ is empty of
vertices.

Let $\Pair{p}{q}$ be a loose pair due to an empty left (resp. right)
leaf, and $c$ be the furthest point from $pq$ on the boundary of the
leaf.  See \figref{crescent} (a).  Let $d$ be the disk centered at $c$
having $p$ and $q$ on its boundary.  We call the region $d \setminus
\Leaf_l(p,q)$ (resp. $d \setminus \Leaf_r(p,q)$) the left (resp.
right) \emph{crescent} of $p q$ and denote it by $\Crescent_l(pq)$
(resp. $\Crescent_r(pq)$).

Crescent of a loose pair $\Pair{p}{q}$ may or may not be empty of all
the other vertices.  In the latter case, the \emph{moonstruck} of $pq$
is the vertex $r$ inside the crescent such that the circumdisk of
$pqr$ is empty of all the other vertices, see \figref{crescent} (b).

\begin{figure}[bt]
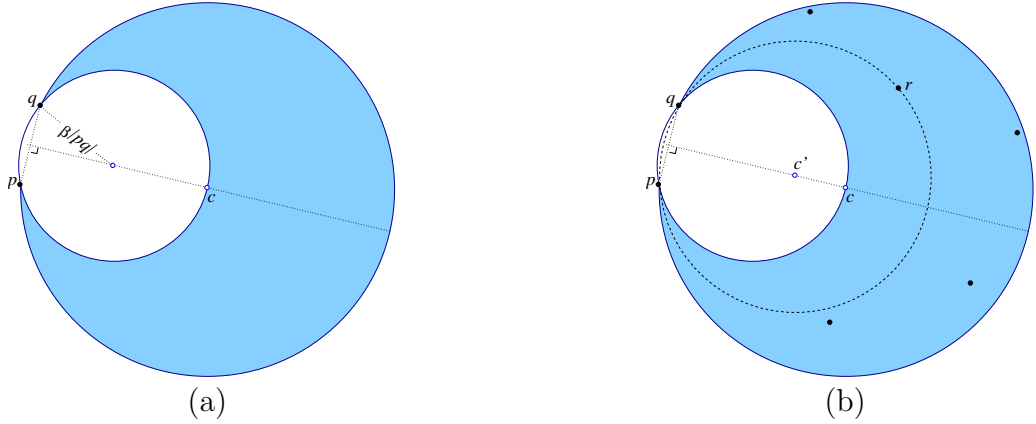

    \TwoFigures{figs/crescent}{width=.65\textwidth}{(a)}
               {figs/moonstruck}{width=.65\textwidth}{(b)}

    \caption{The crescent of a loose pair $\Pair{p}{q}$ is shown as
       the shaded region (a).  If the crescent of $pq$ is empty of all
       the other vertices then the furthest point from $pq$ inside the
       leaf (shown as $c$) is the off-center of $pq$.  Otherwise, the
       moonstruck of a loose pair $\Pair{p}{q}$ with non-empty
       crescent is shown as $r$ (b).  Off-center in this case is the
       circumcenter of $pqr$, shown as $c'$.}
    \figlab{crescent}
\end{figure}

\begin{lemma}
    There exists a loose pair in a point set $P$ if and only if the
    minimum angle in the Delaunay triangulation of $P$ is smaller than
    or equal to $\alpha$.
\end{lemma}

\begin{proof}
    If there exists a loose pair $\Pair{p}{q}$ then $pq$ is a Delaunay
    edge.  Moreover the triangle incident to edge $pq$ on the side of
    the empty leaf must be bad, with an angle smaller than $\alpha$.
    
    For the other direction, let $pqr$ be the bad triangle with the
    shortest edge.  Without loss of generality assume $pqr$ is a
    $\tt{right\_turn}$.  If the right leaf of $pq$ is empty then
    $\Pair{p}{q}$ is a loose pair and we are done.  Otherwise, let $s$
    be the first point a morphing from the circumsphere of $pqr$ to
    right leaf of $pq$ hits (fixing the points $p$ and $q$).  Then
    both $ps$ and $sq$ are shorter features then $pq$ and are incident
    to a bad Delaunay triangle. This is a contradiction to the
    minimality of the $\dist{pq}$.
\end{proof}

This lemma suggest the refinement method depicted in
\algref{sequential:basic}.  Note that each loose pair corresponds to a
bad triangle in the Delaunay triangulation of the growing point set.
Hence, this algorithm is simply another way of stating the Delaunay
refinement with off-centers algorithm presented in \cite{u-ocnts-04}.
\Ungor{} showed that the Delaunay refinement with off-centers
algorithm terminates and the resulting point set is size-optimal.  In
order to give optimal time bounds we will refine this algorithm in the
next section.

\begin{algorithm}[t]
    \caption{\sc Loose Pair Removal} 
    \begin{algorithmic}[l]
        \REQUIRE A point set $\Omega$ in $\Re^2$ and $\beta$%
        \ENSURE A Steiner triangulation of $\Omega$ where all
        triangles have radius-edge ratio at most $\beta$ %
        \STATE Let $P = \Omega$ %
        \WHILE{there exists a loose pair $\Pair{p}{q}$ in $P$}%
        \STATE Insert the off-center of $pq$ into $P$%
        \ENDWHILE%
        \STATE Compute and Output the Delaunay triangulation of the
        resulting point set
    \end{algorithmic}

    \alglab{sequential:basic}
\end{algorithm}

The next two lemmas follow directly from \cite{u-ocnts-04}. They state
that during the refinement process we never introduce new features
that are smaller than the current loose pair being handled.

\begin{lemma}
    Let $P$ be a point-set, $\Pair{p}{q}$ be a loose pair of $P$, and
    $P'$ be the set resulting for inserting the off-center of $pq$. We
    have for any $x\in P$, that if $\lfs_{P'}(x) < \lfs_P(x)$, then
    $\lfs_{P'}(x) \geq \dist{pq}$.  \lemlab{fiction}
\end{lemma}

\begin{lemma}
    Let $\Omega$ be the input point set, and let $P$ be the current
    point set maintained by the refinement algorithm depicted in
    \algref{sequential:basic}. Let $\PFinal$ denote the point set
    generated by \algref{sequential:basic}. Then for any point $p$ in
    the plane, we have throughout the algorithm execution that
    $\lfs_\Omega(p) \geq \lfs_P(p) \geq \lfs_\PFinal(p) \geq
    \ConstLfsShrink \lfs_\Omega(p)$, where $\ConstLfsShrink > 0$ is a
    constant.

    \lemlab{LFS:the:same}
\end{lemma}

\begin{algorithm*}[tb]
    \caption{: \AlgRefine{}} 
    \alglab{QT}
    \begin{algorithmic}[l]
        \REQUIRE A point set $\Omega \subseteq [1/3,2/3]^2 \in \Re^2$ and $\beta$
        \ENSURE A Steiner triangulation of $\Omega$ where all
            triangles have radius-edge ratio at most $\beta$  
        \STATE Split each edge of the square $[0,1]^2$ into three segments.
        \STATE Construct a balanced quadtree $\QT$ of $\Omega$ using
        $[0,1]^2$ for the root node. %
        \STATE Insert all the nodes of $\QT$ into a heap $\Heap$,
        sorted from smallest node to largest.%
        \STATE Initialize all vertices to be active.%
        \STATE Let $prev\_i$ be the depth of the smallest $\QT$ node.
        \WHILE{$\Heap$ is not empty} \STATE $\Cell \leftarrow
        \extractMin(\Heap)$.  \STATE $i = depth(\Cell)$.  \IF{$i <
           prev\_i$}%
        \STATE Move all vertices in level $prev\_i$ which are also
        active in level $i$ to the $i$th level.  \STATE $prev\_i =
        i$.%

        \ENDIF \FOR{every active vertex $p \in P \cap \Cell$}
        
        \FOR{every active vertex $q \in P$ such that $\dist{pq} \le
           \ConstReach \Size(\Cell)$}

        \IF {$pq$ is loose}%
        \STATE Insert the $r = \OffCenter(p,q)$, and store the $r$ in
        $\QT$ in a cell $\CellA$ as low as possible, such that
        $\ConstLower \Size(\CellA) \leq \dist{pr} \leq \ConstUpper
        \Size(\CellA)$ and $\Size(\CellA) \geq \Size(\Cell)$

        \ENDIF  
        \STATE for every node in the same level of $\Cell$ that had a
        point inserted into it, because of the above step,
        reinsert it into the heap $\Heap$.
        \ENDFOR
        \ENDFOR
        \ENDWHILE
        \STATE Compute and Output the Delaunay triangulation of the
        resulting point set
    \end{algorithmic}
\end{algorithm*}

\lemref{fiction} suggests a natural algorithm for generating a good
cloud of points. Since inserting a new off-center can not decrease the
smallest feature of the point cloud, it is natural to first handle the
shortest loose pair first. Namely, repeatedly find the smallest loose
pair, insert its off-center, till there are no loose pairs left.
Because the domain is compact, by a simple packing argument it follows
that this algorithm terminates and generates an optimal mesh.  This is
one possible implementation of (the generic)
\algref{sequential:basic}.

Implementing this in the naive way, is not going to be efficient.
Indeed, first we need to maintain a heap sorted by the lengths of the
loose pairs, which is already too expensive. More importantly,
checking if a pair is loose requires performing local queries on the
geometry which might be too expensive to perform.

We will overcome these two challenges by handling the loose pairs
using a weak ordering on the pairs.  This would be facilitated by
using a quadtree for answering the range searching queries needed for
the loose pairs determination.  In particular, our new algorithm is
just going to be one possible implementation of
\algref{sequential:basic}, and as such \lemref{fiction} and
\lemref{LFS:the:same} hold for it.

\section{Efficient Algorithm Using a Quadtree}
\seclab{new:alg}

We construct a balanced quadtree for $\Omega$, using the unit square
as the root of the quadtree.  In the following, $P$ denotes the
current point set, as it grows during the algorithm execution.  Let
$\PFinal$ be the final point set generated.  A \emph{balanced}
quadtree has the property that two adjacent leaves have the same size
up to factor two.  A balanced quadtree can be constructed, in $O(n
\log n + m)$ time, where $m$ is the size of the output, see
\cite{beg-pgmg-94}.  Such a balanced quadtree also approximates the
local feature size of the input, and its output size $m$ is
(asymptotically) the size of the cloud of points we need to generate.
In the constructed quadtree we maintain, for each node, pointers to
its neighbors in its own level, in the quadtree, and to its neighbors
in the levels immediately adjacent to it.


The new algorithm is depicted in \algref{QT}. For the time being,
consider all points to be active throughout the execution of the
algorithm. Later, we will demonstrate that it is enough to maintain
only very few active points inside each cell, thus resulting in a fast
implementation.  We show that each quadtree node is rescheduled into
the heap at most a constant number of times, implying that the
algorithm terminates. 

In \algref{QT}, collecting the active points around a cell $\Cell$,
checking whether a pair is active, or finding the moonstruck point of
a pair is done by traversing the cells adjacent to the current cell,
using the boundary pointers of the well-balanced quadtree. We will
show that all those operations takes constant time per cell.

One technicality that is omitted from the description of \algref{QT},
is that we refine the boundary edges of the unit square by splitting
such an edge in the middle, if it is being encroached upon.  Since the
local feature size on the boundary of the unit square is $\Theta(1)$,
by \lemref{LFS:the:same}. As such, this can automatically handled
every time we introduce a new point, and it would require $O(1)$ time
for each insertion. This guarantees that no point would be inserted
outside the unit square. Note, that in such a case the encroaching new
vertex is not being inserted into the point set (although it might be
inserted at some later iteration).

\subsection{Proof of Correctness}

The proof of correctness is by induction over the depth of the nodes
being handled. We use $d_{\QT}$ to denote the depth of the
quadtree $\QT$. In the $k$th stage of the execution of the algorithm,
it handles all nodes of depth $(d_{\QT} - k)$ in the tree. Next, the
algorithm handles all nodes of depth $(d_{\QT} - (k + 1))$, and so on.

By the balanced quadtree construction \cite{beg-pgmg-94}, for every
leaf $\Cell$ of the quadtree $\QT$, and every point $p \in
P \cap \Cell$, we have $\ConstLower \Size(\Cell) \leq \lfs_P(p) \leq
\ConstUpper \Size(\Cell)$, where $\ConstLower$ and $\ConstUpper$ are
prespecified constants such that $\ConstUpper \geq 2 \ConstLower$. In
particular, the value of $\ConstUpper$ and $\ConstLower$ is determined
by the initially constructed quadtree.

\begin{lemma}
    Let $P$ be the current point set maintained by \algref{QT}, and
    let $r$ be an off-center of a loose pair $\Pair{p}{q}$ in $P$.
    Let $\CellA$ be the quadtree node that the point $r$ is inserted
    into.  We have $\ConstLower \cdot \Size(\CellA) \leq \lfs_P(r)
    \leq \ConstUpper \cdot \Size(\CellA)$. In particular, $
    \ConstLowerFinal \cdot \Size(\CellA) \leq \lfs_\PFinal(r) \leq
    \ConstUpper \cdot \Size(\CellA)$, where $\ConstLowerFinal =
    \ConstLfsShrink \cdot \ConstLower$.
    
    \lemlab{LFS:qt}
\end{lemma}

\begin{proof}
    The claim follows from the explicit condition used in the
    insertion part of the algorithm. Observe that since $\lfs_P(r) =
    \dist{pr} \geq \lfs_P(p) \geq \ConstLower \cdot \Size(\Cell)$,
    where $\Cell$ is the cell of the quadtree containing the point
    $p$. As such, a node $\CellA$ that contains $r$ and is in the same
    level as $\Cell$, will have $\ConstLower \cdot \Size(\CellA) \leq
    \lfs_P(p) \leq \lfs_P(r)$, by induction. If $\lfs_P(r) \leq
    \ConstUpper \cdot \Size(\CellA)$ then we are done. Otherwise,
    $\lfs_P(r) > \ConstUpper \cdot \Size(\CellA)$ implying that
    $\ConstLower \cdot \Size( \Parent( \CellA)) \leq \ConstUpper \cdot
    \Size(\CellA) \leq \lfs_P(r)$, since $\ConstUpper \geq 2
    \ConstLower$.  Thus, set $\CellA \leftarrow \Parent(\CellA)$ and
    observe that $\ConstLower \cdot \Size(\CellA) \leq \lfs_P(r)$, as
    such we can continue climbing up the quadtree till both
    inequalities hold simultaneously.

    The second part follows immediately from \lemref{LFS:the:same}.
\end{proof}

\begin{lemma}
    Off-center insertion takes $O(1)$ time.
\end{lemma}
\begin{proof}
    Let $r$ be an off-center of a loose pair $\Pair{p}{q}$, and let
    $\Cell$ and $\CellA$ be the cells of the quadtree containing $p$
    and $r$, respectively.  By \lemref{LFS:qt}, $\lfs_P(p) = \Theta(
    \Size(\Cell))$. By the algorithm definition, we have $\dist{pq} =
    \Theta(\Size(\Cell))$. As such, $\dist{pr} = \Theta(\dist{pq}) =
    \Theta( \Size(\Cell))$. Namely, $\lfs_P(r) = \dist{pr} = \Theta(
    \Size(\Cell))$. Namely, in the grid resolution of $\Cell$, the
    points $p$ and $r$ are constant number of cells away form each
    other (although $r$ might be stored a constant number of levels
    above $\Cell$). Since every node in the quadtree have cross
    pointers to its immediate neighbors in its level (or the above
    level), by the well-balanced property of the quadtree.  It follows
    that we can traverse from $\Cell$ to $\CellA$ using constant time.
\end{proof}

\begin{lemma}
    The shortest loose pair of $P$ in the beginning of the $i$th stage
    is of length at least $\eta_i = \ConstLowerFinal /
    2^{d_{\QT}-i+1}$.

    \lemlab{inductive}
\end{lemma}

\begin{proof}
    For $i=1$, the claim trivially holds by the construction of the
    balanced quadtree using \cite{beg-pgmg-94}.
    Now, assume that the claim holds for $i=1,
    \ldots, k$. We next show that the claim holds for $i=k+1$.
    Specifically, it holds at the end of the $k$th stage.
    
    Suppose for the sake of contradiction that there exists a loose
    pair $\Pair{p}{q}$ shorter than $\eta_k$. Assume, without lose of
    generality, that $p$ was created after $q$ by the algorithm, and
    let $\Cell$ be the node of $\QT$ that contains $p$.  If the depth
    of $\Cell$ is $d_{\QT}-k$, then since $\dist{pq} \leq \eta_k \leq
    \ConstReach \Size(\Cell)$, we have that the algorithm handled the
    point $p$ in $\Cell$, it also considered the pair $\Pair{p}{q}$
    and inserted its off-center. So, the pair $\Pair{p}{q}$ is not
    loose at the end of the $k$th stage.

    If the depth of $\Cell$ is larger then $d_{\QT}-k$ then $\lfs(p)$
    was larger than $\ConstLower \Size(\Cell)$ when $p$ was inserted.
    As such, $\lfs_P(p) > \ConstLowerFinal \Size(\Cell) \geq \eta_k$,
    which is a contradiction, since any loose pair that $p$
    participates in must be of length at least $\lfs_P(p)$.
\end{proof}

Note that when the algorithm handles the root node in the last
iteration, it ``deteriorates'' into being \algref{sequential:basic}
executed on the whole point set.  Hence \lemref{inductive} implies the
following claim.

\begin{claim}
    In the end of the execution of \algref{QT}, there are no loose
    pairs left in $\PFinal$, where $\PFinal$ is the point set
    generated by the algorithm.
\end{claim}

\subsection{How the refinement evolves}

Our next task, is to understand how the refinement takes place around
a point, and form a ``protection''area around it. In particular, the
region around a point $p \in P$ with $\lfs_\Omega(p)$ is going to be
effected (i.e., points inserted into it), starting when the algorithm
handles cells of level $i$, where $1/2^i \approx \lfs_\Omega(p)$.
Namely, the region around $p$ might be refined in the next few levels.
However, after a constant number of such levels, the point $p$ is
surrounded by other points, and $p$ is not loose with any of those
points. As such, the point $p$ is no longer a candidate to be in a
loose pair. To capture this intuition, we prove that this encirclement
process indeed takes place.

We define the \emph{gap} of a vertex $x \in P$ (which is not a
boundary vertex), denoted by $\gap(x)$, as the ratio between the
radius of the largest disk that touches $x$ and does not contain any
vertex inside, and the $\lfs_P(x)$.

\begin{lemma}
    For a vertex $w \in P$, if $\gap(w) > \ConstGap =
    (2\beta)^{\pi/\alpha+1}$, then there exists a loose pair of
    $P$ of length $\leq \ConstGapBlowup \cdot\lfs(w)$, where
    $\ConstGapBlowup = (2\beta)^{\pi/\alpha}$.

    \lemlab{gap}
\end{lemma}

\begin{proof}
    Assume that $w$ is strictly inside the bounding square $B$. Proof
    for the case where $w$ is on the boundary of $B$ is similar and
    hence omitted.  Let $T_1, T_2, \ldots, T_m$ be the Delaunay
    triangles incident to $w$ and $u_1, u_2, \ldots, u_m$ be the
    Delaunay neighbors of $w$.  Note that if $\dist{w u_i} \geq 2
    \beta \dist{w u_{i-1}}$, then the left flower of $w u_{i-1}$ must
    be empty and hence $w u_{i-1}$ is a loose pair.  Similarly, if
    $\angle{w u_i u_{i-1}} < \alpha$ then $\Pair{w}{ u_{i-1}}$ is a
    loose pair.  If $\angle{w u_i u_{i-1}} \geq \alpha$ and
    $\angle{u_i w u_{i-1}} < \alpha$ then by the law of sines, it must
    be that $\dist{u_i u_{i-1}} \leq \dist{u_{i-1}w}$, and $u_i
    u_{i-1}$ is facing an angle smaller than $\alpha$, and as such it
    is a loose pair.

    Suppose that none of the triangles $T_1,\ldots, T_m$ have a loose
    pair on their boundary. Then, it must be that $m \leq
    2\pi/\alpha$, since the angle $\angle u_{i} w u_{i+1} \geq
    \alpha$, for $i=1,\ldots, m$. But then, $\dist{u_i w} \leq
    (2\beta)^{i-1} \lfs(w)$ and $\dist{u_i w} \leq (2\beta)^{m-i+1}
    \lfs(w)$. It follows that $\dist{u_i w} \leq (2\beta)^{\pi/\alpha}
    \lfs(w)$, for $i=1,\ldots, m$. Since, all the angles in $T_i$ are
    larger than $\alpha$, it follows that circumcircle of $T_i$ is of
    radius $\leq \beta \dist{u_i w} \leq
    \beta(2\beta)^{\pi/\alpha}\lfs(w)$. But then, the gap around $w$,
    is at most $\beta(2\beta)^{\pi/\alpha}$. A contradiction, since
    $\gap(w) = \ConstGap > \beta(2\beta)^{\pi/\alpha}$.
    
    Thus, one $T_1,\ldots, T_m$ must be bad. Arguing as above, one can
    show that the first such triangle, has a loose pair of length $\leq
    (2\beta)^{\pi/\alpha} \lfs(w)$, as claimed.
\end{proof}

\lemref{gap} implies that if we handle all loose pairs of length
smaller than $\ell$, then all the points having a big gap,
must be with local feature size $\Omega(\ell)$.

\begin{lemma}
    Let $P$ be a point set such that all the loose pairs are of length
    at least $\ell/\ConstA$, for a constant $\ConstA \geq 2$.  Let
    $\Pair{p}{q}$ be a loose pair of length $\ell$ with a non-empty
    crescent, and let $w$ be its moonstruck neighbor.  Then, $\lfs(w)
    \geq \frac{\ell}{\ConstA \ConstGapBlowup}$.

    \lemlab{moonstruck}
\end{lemma}

\begin{proof}
    Since $w$ is a moonstruck point of a loose pair of length at least
    $\ell$ there is an empty ball of radius at least $\beta \ell$
    touching $w$. Hence, $\gap(w) \geq \frac{\beta \ell}{\lfs(w)}$.
    We consider two cases.  If $\frac{\beta \ell}{\lfs(w)} \geq
    \ConstGap$, then by \lemref{gap}, there exist a loose pair of size
    at most $\ConstGapBlowup \lfs(w)$.  However, all loose pairs are
    of length $\geq \ell/\ConstA$, and it follows that
    $\ConstGapBlowup \lfs(w) \geq \ell/\ConstA$.  Hence, $\lfs(w) \geq
    \ell/(\ConstA \ConstGapBlowup)$.  On the other hand, if
    $\frac{\beta \ell}{\lfs(w)} \leq \ConstGap$ then,
    \[
    \lfs(w) \geq \frac{\beta \ell}{\ConstGap}
     \geq \frac{ \ell}{\ConstGap}
      \geq \frac{ \ell}{\ConstGapBlowup}
    \geq \frac{\ell}{\ConstA \ConstGapBlowup},
    \]  
    since $\ConstGapBlowup \geq \ConstGap$ and $\beta \geq 1$.
\end{proof}

\subsection{Managing Active Points}

As we progress with execution of the algorithm, the results of the
previous section imply that a vertex with relatively small feature
size cannot participate in a loose pair, nor be a moonstruck point.
So, in the evolving quadtree we do not maintain such set of points
that play no role in the later stages of the algorithm execution.
This facilitates an efficient search for finding the loose pairs and
moonstruck points as shown in the rest of this section.  For each
vertex, size of its insertion cell gives a good approximation of its
feature size. We use this to determine the lifetime of each vertex in
our evolving quadtree data structure.

The \emph{activation depth} of an input vertex $p$, denoted by
$\ACT{p}$, is the level of the initial quadtree leaf containing $p$.
For a Steiner point $p$, the \emph{activation depth} is the depth
of the cell $p$ is inserted into.

\begin{lemma}
    A vertex $p$ can not be a loose pair end or a moonstruck point,
    when the algorithm handles level of depth $< \ACT{p} -
    \ConstLifeSpan$, where $\ConstLifeSpan = \lg \frac{\ConstGapBlowup
       \ConstUpper}{\ConstLowerFinal} + 1$.

    \lemlab{life:span}
\end{lemma}

\begin{proof}
    When the point $p$ was created, we had $\lfs_P(p) \leq \ConstUpper
    / 2^\ACT{p}$, by \lemref{LFS:qt}. Now, if $p$ is an endpoint of a
    loose pair in depth $m\leq \ACT{p}$ in the quadtree, it must be
    that the length $\ell$ of this pair is at least $\ConstLowerFinal
    / 2^{m}$, by \lemref{inductive}. Since the local feature size
    $\lfs_P$ is a non-increasing function as our algorithm progresses,
    it follows that
    \[
    \gap(p) \geq \frac{\ell}{\lfs_P(p)} \geq \frac{\ConstLowerFinal /
       2^{m}}{\ConstUpper /2^\ACT{p}} = 2^{\ACT{p}-m}
    \cdot \frac{\ConstLowerFinal}{\ConstUpper}.
    \]
    If $\gap(p) \leq \ConstGap$ then $2^{\ACT{p}-m} \cdot
    \frac{\ConstLowerFinal}{\ConstUpper} \leq \ConstGap$. Implying
    that $\ACT{p}-m \leq \lg \frac{\ConstGap
       \ConstUpper}{\ConstLower}$.
    
    By \lemref{gap}, if $\gap(p)$ at any point in the algorithm
    becomes larger than $\ConstGap$, then there exists a loose pair of
    length $\leq \ConstGapBlowup \cdot\lfs_P(p)$. But all such pairs
    are handled in level $ \geq \LevelInd$ in the quadtree, where
    $\ConstGapBlowup \cdot\lfs_P(p) \geq \ConstLowerFinal /
    2^{\LevelInd+1}$ by \lemref{inductive}. Thus, $\ConstGapBlowup
    \ConstUpper /2^\ACT{p} \geq \ConstGapBlowup \lfs_P(p) \geq
    \ConstLowerFinal / 2^{\LevelInd+1}$. Implying that
    \[
    \LevelInd \geq \rho = \ACT{p} - \lg \frac{ \ConstGapBlowup
       \ConstUpper}{\ConstLowerFinal} - 1.
    \]    
    This implies, that when the algorithm handles cells of depth $1,
    \ldots, \rho$, we have that the vertex $p$ can not participate
    directly in a loose pair.

    If $p$ is not loose pair end, but is a moonstruck point 
       for a loose pair, then 
    \[
    \frac{\ConstUpper}{2^\ACT{p}} \geq \lfs_P(p) \geq
    \frac{\ConstLowerFinal}{2^{\LevelInd+1} \ConstGapBlowup}
    \]
    by \lemref{moonstruck} and \lemref{inductive}. This in turn
    implies that $\LevelInd \geq  \ACT{p} - \lg \frac{ \ConstGapBlowup
       \ConstUpper}{\ConstLowerFinal} - 1$.
\end{proof}

\begin{defn}
    A point $p$ is \emph{active} at depth $i$, if $\ACT{p} \geq i \geq
    \ACT{p} - \ConstLifeSpan$, where $\ConstLifeSpan$ is a constant
    specified in \lemref{life:span}.
\end{defn}

Note, that the algorithm can easily maintain the set of the active
points.  \lemref{life:span} implies that only active points are needed
to be considered in the loose pair computation.

\begin{observation}
    During the off-center insertion any new loose pairs introduced are
    at least the size of the existing loose pairs.
\end{observation}

\begin{lemma}
    At any stage $i$, the number of active points inside a cell at
    level $d_\QT - i$ is a constant.
\end{lemma}

\begin{proof}
    This is trivially true in the beginning of the execution of the
    algorithm, as the initial balanced quadtree has at most a constant
    number of vertices in each leaf. Later on, \lemref{life:span}
    implies that when a point $p$ is being created, with
    $\ell=\lfs(p)$, then its final local feature size is going to be
    $\Theta(\ell)$. To see that, observe that when $p$ was created,
    the algorithm handled loose pairs of size $\Omega(\ell)$. From
    this point on, the algorithm only handle loose pairs that are
    longer (or slightly shorter) than $\ell$.  Such a loose pair, can
    not decrease the local feature size to be much smaller than
    $\ell$, by \lemref{fiction}.
    
    This implies that when $p$ is being created, we can place around
    it a ball of radius $\Omega(\lfs(p))$ which would contain only $p$
    in the final generated point set. Since $p$ becomes inactive
    $\ConstLifeSpan$ levels above the level it is being created, it
    follows that a call in the quadtree can contain at most a constant
    number of such protecting balls, by a simple packing argument.
\end{proof}

\subsection{Efficient Implementation Details}

The above discussion implies that during the algorithm execution, we
can maintain for every quadtree node a list of constant size that
contains all the active vertices inside it. When processing a node, we
need to extract all the active points close to this cell $\Cell$. This
requires collecting all the cells in this level, which are constant
number of cells away from $\Cell$ in this grid resolution. In fact,
the algorithm would do this point collection also in a constant number
of levels above the current level, so that it collects all the Steiner
points that might have been inserted. Since throughout the execution
of the algorithm we maintain a balanced quadtree, we have from every
node, pointers to its neighbors either in its level, or at most one
level up. As such, we can collect all the neighbors of $\Cell$ in
constant distance from it in the quadtree, in constant time, and
furthermore, extract their active points in constant time. Hence,
handling a node in the main loop of \algref{QT} takes constant time.

We need also to implement the heap used by the algorithm.  We store
nodes in the heap $\Heap$, and it extracts them according to their
depth in the quadtree. As such, we can implement it by having a
separate heap for each level of the quadtree.  Note that the local
feature size of a vertex when inserted into a quadtree node is within
a constant factor of the size of the node.  Hence a node can be
rescheduled in the heap at most a constant number of times.  For each
level, the heap is implemented by using a linked list and a
hash-table. Thus, every heap operation takes constant time.

\subsection{Connecting the Dots}

We shall also address how to perform the final step of \algref{QT},
that is computing the Delaunay triangulation of the resulting point
set $\PFinal$.  This can be done by re-executing a variant of the main
loop of \algref{QT} on $\PFinal$, which instead of refining the point
set, reports the Delaunay triangles.  We use a similar deactivation
scheme to ignore vertices whose all Delaunay triangles are reported.
Since $\PFinal$ is a well-spaced point set, for a pair of nearby
active vertices we can efficiently compute whether the two makes a
Delaunay edge and if so locate also the third point that would make
the Delaunay triangle.  It is straightforward but tedious to argue
that the running time of this algorithm is going to be proportional to
the running time of \algref{QT}.

\subsection{Analysis}

The initial balanced quadtree construction takes $O(n \log n + m)$
time, where $m$ is the size of the resulting quadtree \cite{beg-pgmg-94}. 
This quadtree has the property that the size length of a leaf is
proportional to the local feature size. This in turn implies the value
of $\ConstLower$ and $\ConstUpper$, which in turn guarantees that no
new leafs would be added to the quadtree during the refinement process.

Furthermore, the point set generated by \algref{QT} has the property
that its density is proportional to the local feature size of the
input. Namely, the size of the generated point-set is $O(m)$.  Since
all the operations inside the loop of \algref{QT} takes constant time,
we can charge them to either the newly created points, or to the
relevant nodes in the quadtree. This immediately implies that once the
quadtree is constructed, the running time of the algorithm is $O( m
)$.

\begin{theorem}
    Given a set $\Omega$ of $n$ points in the plane the Delaunay
    refinement algorithm (depicted in \algref{QT}) computes a
    quality-guaranteed size-optimal Steiner triangulation of $\Omega$,
    in optimal time $O(n \log n + m)$, where $m$ is the size of the
    resulting triangulation.
\end{theorem}

\section{Conclusions}
\seclab{conclusions}

We presented a time-optimal algorithm for Delaunay refinement in the
plane. It is important to note that the output of this new algorithm
is the same as that of the off-center based Delaunay refinement
algorithm given in \cite{u-ocnts-04}, which \emph{outperforms} the
circumcenter based refinement algorithms in practice.  The natural
open question for further research is extending the algorithm in three
(and higher) dimensions. We believe that extending our algorithm to
handle PSLG in the plane is doable (with the same time bounds), but is
not trivial, and it would be included in the full version of this
paper.

We note that when building the initial quadtree, we do not have to
perform as many refinement steps as used in the standard quadtree
refinement algorithm of Bern \etal{} \cite{beg-pgmg-94}. While their
algorithm considers a quadtree cell with two input vertices crowded
and splits it into four, we are perfectly satisfied with a balanced
quadtree as long as the quadtree approximates the local feature size
within a constant and hence the number of features in a cell is
bounded by a constant. This difference in the depth of the quadtree
should be exploited for an efficient implementation of our algorithm
(this effects the values of the constants $\ConstLower$ and
$\ConstUpper$).

Parallelization of quadtree based methods are well understood
\cite{bet-pcqqt-99}, while design of a theoretically optimal and
practical parallel Delaunay refinement algorithm is an ongoing
research topic \cite{stu-pdraa-02,stu-tcppspia-04}.  We believe our
approach of combining the strengths of quadtrees as a domain
decomposition scheme and Delaunay refinement with off-centers will
lead to a good parallel solution for the meshing problem.


 
\bibliographystyle{salpha} 
\bibliography{shortcuts,geometry}

\begin{table*}[t]
    \centerline{
       \begin{tabular}{|l||l|l|}
           \hline
           Notation & Value & Comment\\
           \hline \hline 
           $\Omega$ & & Input point set.\\
           \hline
           $P$      & & Current point set maintained by the
           algorithm.\\
           \hline
           $\PFinal$ & & Final point set\\
           \hline
           $\beta$ & $\geq \sqrt{2}$ & radius-edge ratio threshold for
                                       bad triangles\\
           \hline
           $\alpha$ & $\arcsin( 1/2\beta )$ & small angle threshold
           bad triangles\\
           \hline
           $\ConstGap$ & 
           $(2\beta)^{\pi/\alpha+1}$ & Vertex with larger gap than
           $\ConstGap$ participates in a loose pair.\\
           $\ConstGapBlowup$ & $(2\beta)^{\pi/\alpha}$ & Blowup of \lfs{}
           for a lose pair around a vertex with large gap.\\
           $\ConstLower$ &  & Lower bound on the $\lfs$
           of a point inside a leaf of the initial quadtree.\\
           $\ConstUpper$ & $\geq 2 \ConstLower$ &
           Upper bound on the $\lfs$ of a point
           inside a leaf of the quadtree.\\ 
           \hline
           $\ConstReach$ & $\ConstReachExpr$ & 
           \begin{minipage}{0.58\linewidth}
               ~\\[-0.1cm]
               $\ConstReach \Size(\Cell)$ is an upper bound on the
               length of a (relevant) lose pair involving a point $p$
               stored in a cell $\Cell$. For correctness, it required
               that $\ConstReach \geq 2 \sqrt{2}$.
           \end{minipage}\\
           \hline
           $\ConstLfsShrink$ & $>0$ &
           For any point $x$, we have $\lfs_\PFinal(x) \geq
           \ConstLfsShrink \cdot \lfs_\Omega(x)$. \\           
           \hline
           $\ConstLowerFinal$ & $\ConstLower \cdot \ConstLfsShrink$ & 
           \begin{minipage}{0.58\linewidth}
               ~\\[-0.1cm]               
               Lower bound on the $\lfs$ of a point $p$ stored inside
               a node $\Cell$ of the quadtree through the algorithm
               execution. Namely, $\lfs_\PFinal(p) \geq
               \ConstLowerFinal \Size( \Cell)$.
           \end{minipage}\\
           \hline
       \end{tabular}
    }
    \caption{Notation used in the paper.}
    \tablab{notations}
\end{table*}

\end{document}